\documentclass[final, runningheads]{llncs}
\usepackage[T1]{fontenc}
\usepackage{graphicx}
\usepackage{hyperref}
\usepackage{color}

\usepackage{amsmath}
\newcommand{\itadata}{\footnotesize \textsl{Workshop Scientific HPC in the pre-Exascale era (part of ITADATA2024)}}
\usepackage{fancyhdr}
\fancypagestyle{empty}{\fancyhf{}\fancyhead[C]{\itadata}
  \fancyhead[R]{\includegraphics[width=.05\textwidth]{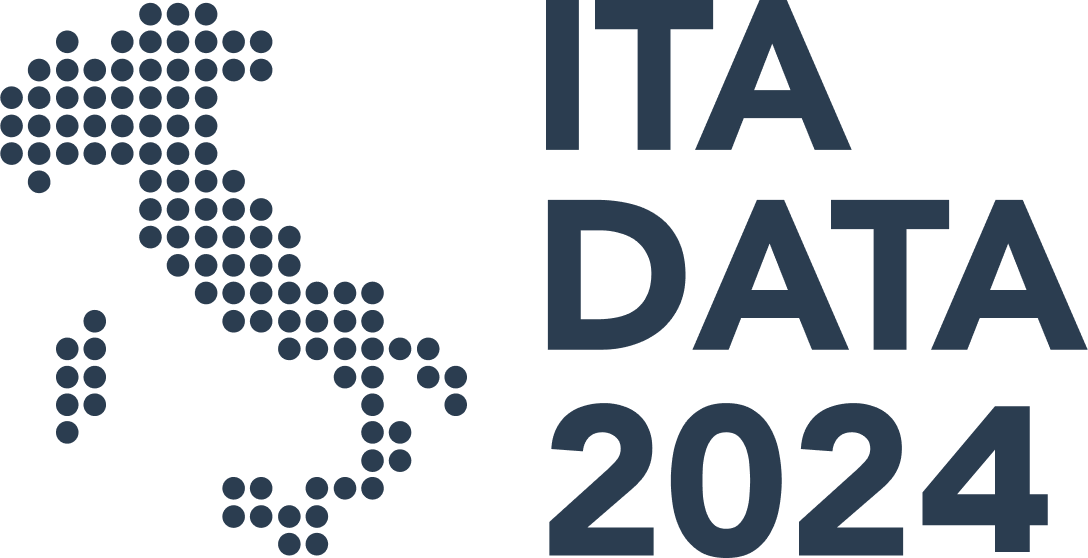}}}

\makeatletter
\begin{document}
\title{From Local to Remote: VisIVO Visual Analytics in the Era of the Square Kilometre Array}
\author{Giuseppe Tudisco\inst{1}\orcidID{0000-0002-2235-3291} \and
Fabio Vitello\inst{1}\orcidID{0000-0003-2203-3797} \and
Eva Sciacca\inst{1}\orcidID{0000-0002-5574-2787} \and
Ugo Becciani\inst{1}\orcidID{0000-0002-4389-8688}}
\authorrunning{G. Tudisco et al.}
\institute{INAF - Osservatorio Astrofisico di Catania, 95123 Catania, Italy}
\maketitle              \begin{abstract}
The field of astrophysics is continuously advancing, with an ever-growing influx of data requiring robust and efficient analysis tools. As the Square Kilometre Array (SKA) radio telescopes come fully operational, we anticipate the generation of hundreds of petabytes of data annually, characterized by unprecedented resolution and detail. In this context, scientific visualization becomes a critical component, enabling researchers to interpret complex datasets and extract meaningful insights. The immense volume of data demands not only suitable tools but also substantial infrastructure and computational capacity to analyze it effectively. In this work, we will discuss how we are addressing these challenges with the development of our interactive visualization tool named VisIVO Visual Analytics. The tool is transitioning from a local visualizer to a remote visualizer, utilizing a client-server architecture. This evolution will allow the software to run parallel visualization pipelines on high-performance computing (HPC) clusters, thereby enhancing its capacity to handle extensive datasets efficiently.

\end{abstract}
\section{Introduction}
Scientific visualization plays a pivotal role in translating complex datasets into meaningful insights, helping researchers comprehend intricate astrophysical phenomena. The main goal of these visualization techniques is to provide tangible outputs, such as images, 3D models, and animations, which allow researchers to explore subtle properties and uncover relationships hidden within vast and complex datasets. Beyond mere representation, advanced visualization tools enable in-depth investigations of aspects that would otherwise be challenging to analyze.

As astrophysical datasets are growing in both scale and complexity -- particularly with the forthcoming Square Kilometre Array (SKA) expected to generate a staggering 700 petabytes of processed data annually -- the development of more sophisticated visualization tools becomes essential for managing and analyzing such massive datasets. The sheer volume and complexity of data produced by state-of-the-art astronomical instruments necessitate the creation of advanced algorithms, computational techniques, and storage solutions capable of handling these ever-increasing demands.

\section{VisIVO Visual Analytics overview}
Since 2005, the INAF Astrophysical Observatory of Catania has been developing and maintaining the Visualization Interface for the Virtual Observatory (VisIVO), a set of tools designed to facilitate 3D and multidimensional data analysis and visualization. VisIVO enables researchers to uncover previously unknown relationships within complex, multivariate astrophysical datasets, supporting advanced knowledge discovery.

\begin{figure}
    \centering
    \includegraphics[width=0.8\textwidth]{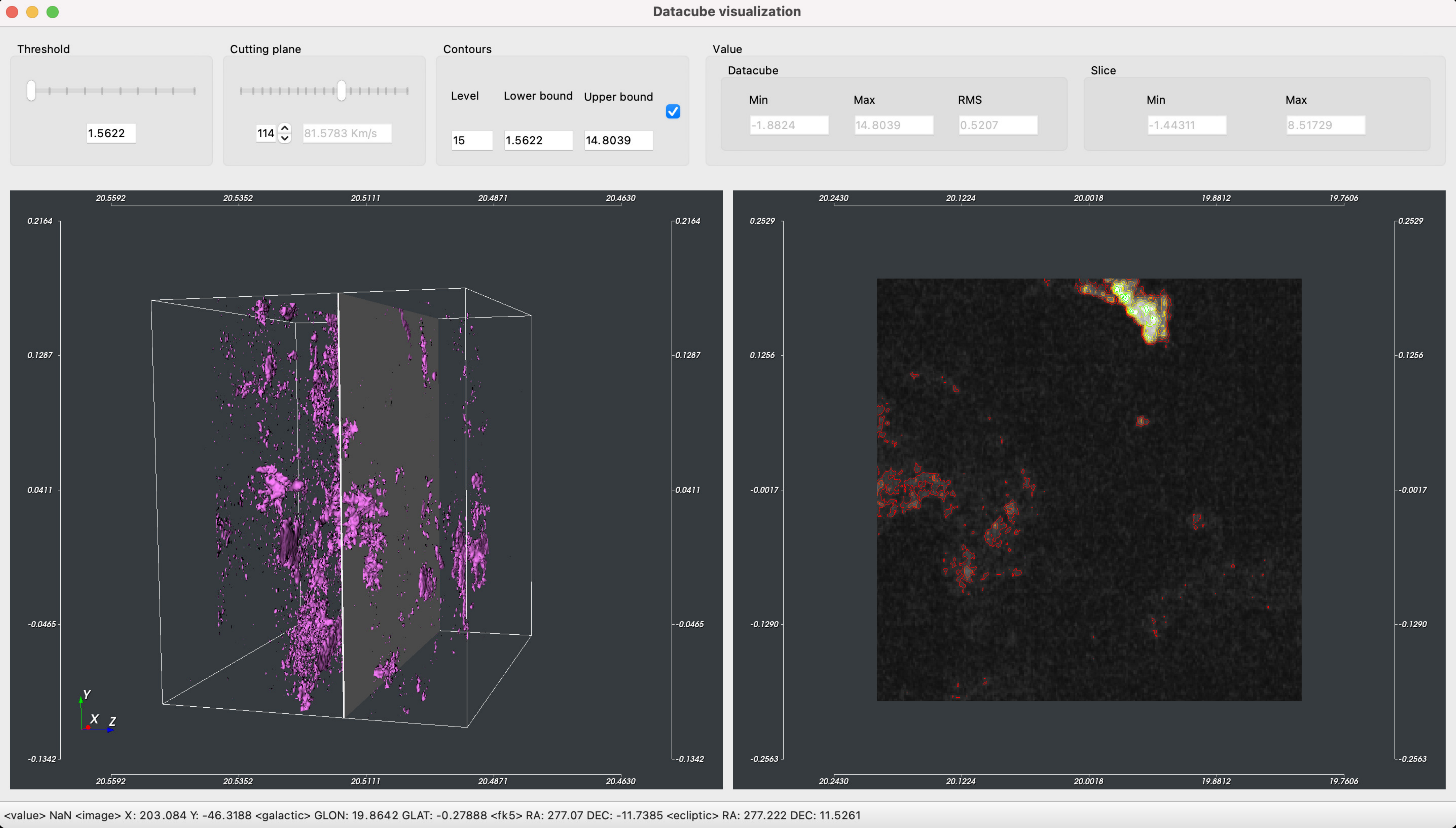}
    \caption{Visualization of a datacube in VisIVO Visual Analytics. The volume rendering is shown on the left, the selected slice is shown on the right.}
    \label{fig:screenshot}
\end{figure}

VisIVO Visual Analytics \cite{Vitello2018}, part of the VisIVO family, is a powerful tool designed to provide an interactive environment for visualizing and analyzing FITS images and cubes. Through its 3D datacube representation, the Visual Analytics tool allows users to systematically explore complex datasets, offering a comprehensive view of data distribution (see Figure \ref{fig:screenshot}). This capability aids in detecting structures and gradients within the three-dimensional space, making it a useful tool for investigating intricate astrophysical phenomena.

VisIVO also offers a variety of analytical tools for detailed exploration of datacubes, including the generation of Position-Velocity diagrams, spectral axis analysis, and calculation of moment maps. These features enable scientists to extract critical information from datacubes, such as velocity profiles, intensity distributions, and velocity dispersions. Beyond handling local files, VisIVO Visual Analytics can search and retrieve the extensive resources of the ViaLactea Knowledge Base \cite{adass_vlkb} through dedicated Search and Cutout services, allowing users to access over 40,000 datasets organized into nearly 100 collections, with a Table Access Protocol service (TAP) providing access to approximately 15 source catalogs.

Users can easily interact with the TAP service through the tool’s intuitive interface, which allows them to define a rectangular image area to specify the region of interest and execute queries for compact sources and filamentary structure catalogs. Additionally, researchers can perform Spectral Energy Distribution (SED) analysis and fitting operations on compact sources overlaid on images. This broad functionality enhances the tool’s versatility, enabling researchers to incorporate external datasets and resources, ultimately expanding the potential for scientific inquiry and analysis.

\subsection{Remote Visualization in VisIVO}
To meet the unique challenges posed by the Square Kilometre Array, efforts are underway to design and implement a remote visualization model within the VisIVO Visual Analytics for visualizing data stored on a separate computational system from the one used for image rendering. Remote visualization for large astrophysical datasets provides several key benefits. This approach not only improves efficiency but also broadens accessibility. By separating data processing from visualization, it optimizes the use of computational resources and overcomes the limitations of common personal hardware.

This approach is being integrated into VisIVO to address the difficulties of managing, accessing, and analyzing massive astrophysical datasets, such as the ones expected from the SKA Observatory. The goal is to develop innovative solutions that enhance both data management and analytical capabilities, ensuring efficient handling of these large-scale datasets. Over the past few years, we have explored and implemented various remote visualization solutions to address this aspect.

\subsubsection{ViaLactea Web}
Our initial approach was ViaLactea Web \cite{adass_vlw}, a web application that offered basic VisIVO Visual Analytics functionality, such as 2D images and 3D cubes visualization (stored on the server), isocontour rendering, and the ability to search and download cutouts from the ViaLactea Knowledge Base. The frontend was built with Node.js, while the backend was developed in Python, with some visualization pipeline code written in C++. These components communicated with each other via web sockets. However, this solution required extensive rewriting of the original VisIVO Visual Analytics codebase to be adapted in a web environment.

\begin{figure}
    \centering
    \includegraphics[width=0.8\textwidth]{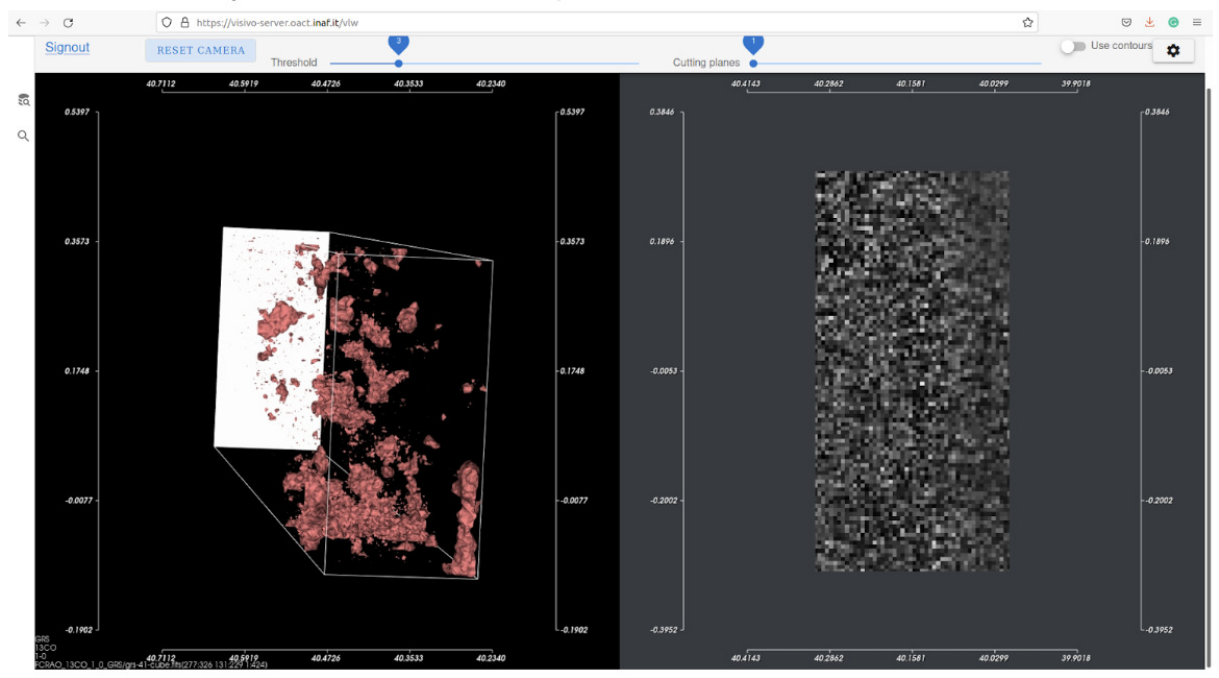}
    \caption{ViaLactea Web running on a server and visualizing a datacube.}
    \label{fig:vlw}
\end{figure}

\subsubsection{Containerization of VisIVO in a VNC environment}
To preserve the full functionality of the existing application while enabling remote data visualization, we developed a VisIVO Visual Analytics container based on an Ubuntu image with a Desktop Environment installed. Thanks to the NVIDIA container toolkit\footnote{\url{https://docs.nvidia.com/datacenter/cloud-native/container-toolkit/}}, TurboVNC\footnote{\url{https://turbovnc.org}}, and VirtualGL\footnote{\url{https://virtualgl.org}}, the container can be deployed on a remote server with GPU support, allowing for hardware-accelerated and high-performance visualization. With appropriate volume mounting, this solution enables users to access and analyze remote datasets through a browser, offering the complete feature set of VisIVO Visual Analytics. Unlike ViaLactea Web, this container-based approach allows users to interact with remote data seamlessly while maintaining the tool’s full capabilities.

\subsubsection{Science Platform}
The next advancement was deploying VisIVO Visual Analytics through Science Platforms, specifically for the CANFAR Science Platform infrastructure and architecture as part of the work developed for the SKA Regional Centres (SRC). The CANFAR Science Platform runs on top of a Kubernetes cluster and allows its users to run application containers (desktop applications, web applications or Jupyter Notebook) in interactive user sessions.

\begin{figure}
    \centering
    \includegraphics[width=0.8\linewidth]{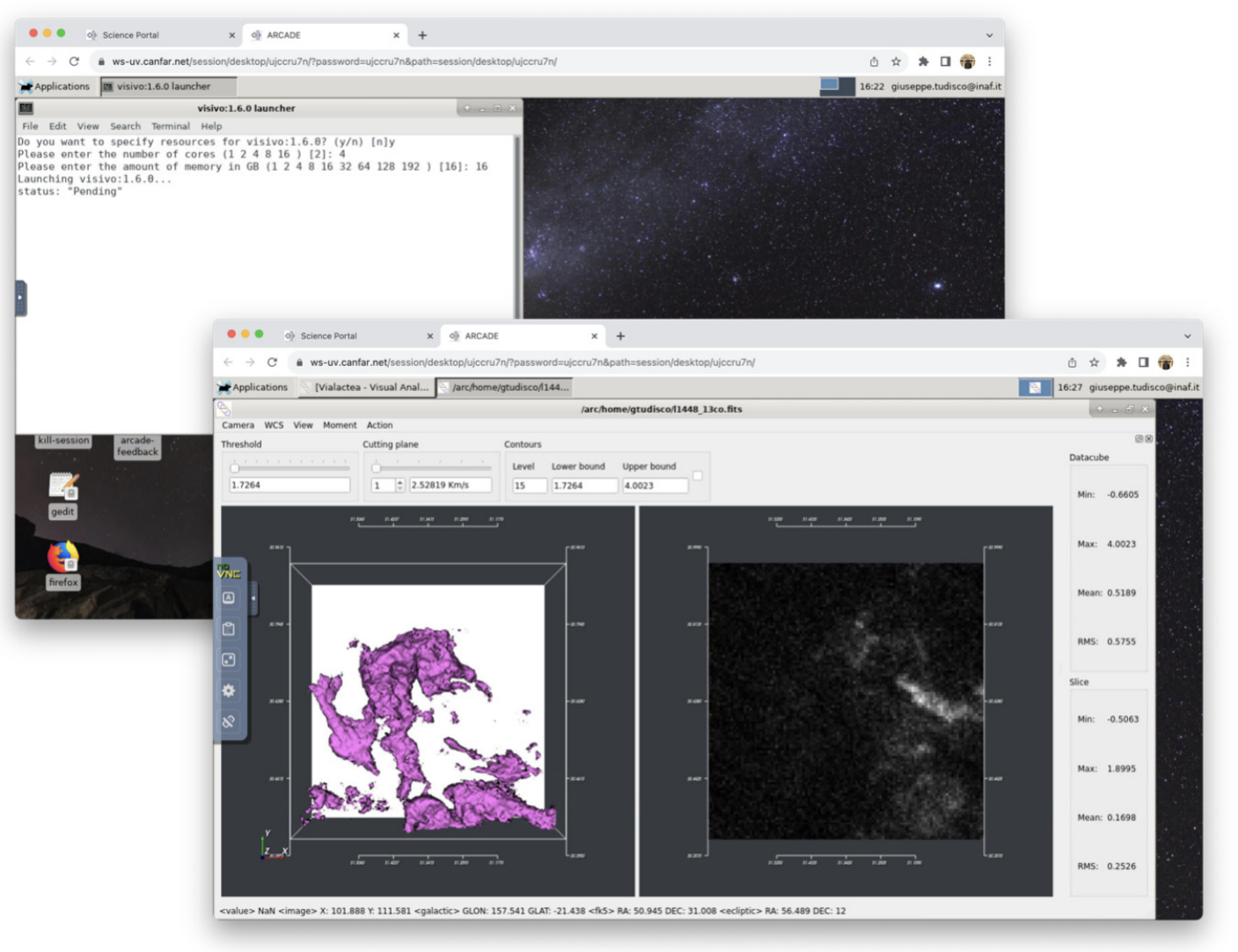}
    \caption{VisIVO Visual Analytics running on a Desktop User Session in CANFAR.}
    \label{fig:canfar}
\end{figure}

We published a VisIVO container image adapted for running on the CANFAR Science Platform (see Figure \ref{fig:canfar}). The desktop sessions are made accessible through noVNC. Users can customize their sessions by specifying the amount of memory and CPU resources needed for the application. All sessions share a common storage system, ensuring persistence of data and settings across sessions, which enhances both the flexibility and scalability of the tool within the Science Platform. This version of VisIVO is going to be available in all the SRC that will deploy CANFAR as part of their services.

\subsubsection{ParaView-based Application}
Finally, our last developments currently involve a new version of VisIVO Visual Analytics that is being developed and is based on ParaView \cite{ahrens200536}. It will incorporate both remote and parallel visualization capabilities through an advanced, distributed architecture. ParaView-based applications consist of three key components: a client application for user interaction, a data server for reading and processing data, and a render server responsible for generating geometries and images, which are then transmitted to the client interface.

For parallel visualization, VisIVO leverages a distributed server architecture running across multiple nodes in an MPI (Message Passing Interface) cluster.
In this scenario, each MPI process works only on a portion of the input datasets to produce partial results that will be combined with the others before being sent to the client.
This enables the computational load of data reading and analysis to be divided among multiple nodes, significantly improving efficiency and throughput and ensuring that large datasets can be processed and visualized seamlessly, even in demanding research environments.

To enhance communication between the client and server in our infrastructure, we are developing a middleware service that manages the lifecycle of server instances via HTTP requests. This service is also designed to handle auxiliary tasks that are not directly related to the visualization pipelines, such as requesting cutouts from the VLKB and storing them remotely. Operating on a designated TCP port, the middleware offers essential server management functions. For example, clients can initiate a server instance using an HTTP POST request, specifying the desired TCP port for the server and, if needed, enabling parallel execution through MPI. Additionally, the service provides endpoints to retrieve execution logs for active server sessions. A summary of the currently available routes is presented in Table \ref{tab:manager}.

\begin{table}[ht]
\centering
\begin{tabular}{|l|l|p{8cm}|}
\hline
\textbf{HTTP Method} & \textbf{Route}       & \textbf{Description}                                                                                                       \\ \hline
GET                  & /info                & Retrieves server information and tells clients if an MPI runtime has been detected.     \\ \hline
POST                 & /server              & Starts a new server instance. The client specifies the port and, if allowed, the number of MPI processes for parallel execution.   \\ \hline
GET                  & /server              & Get info on a running server instance such as start time, process status, and working directory.   \\ \hline
GET                  & /logs                & Retrieves the logs of the currently running server instance, providing insights into execution details and errors (if any).         \\ \hline
\end{tabular}
\caption{VisIVO Server Manager API routes.}
\label{tab:manager}
\end{table}

\section{Conclusion and Future Developments}
In conclusion, the evolution of VisIVO Visual Analytics through the integration of remote and parallel visualization represents a major advancement in addressing the growing demands of modern astrophysical research. As astronomical datasets continue to increase in scale and complexity, transitioning from a desktop-centric tool to a client-server architecture has proven essential for optimizing computational resources and enhancing data accessibility.

We are currently in the process of transitioning our tool's functionalities from a desktop-centric architecture to a more dynamic client-server paradigm. Although the comprehensive porting process is still ongoing, the initial results suggest we are on a positive trajectory in achieving our objectives. The transition to a client-server architecture reflects a deliberate response to the changing needs of modern astrophysics research. As we continue through this process, we are committed to enhancing and perfecting our tool, solidifying its function as a reliable and efficient resource for scientific discovery.

\begin{credits}
\subsubsection{\ackname} The work has received funding from the European Research Council under ERC Synergy Grant ECOGAL (grant 855130), by INAF Mini-Grants RSN5 2023 and by the Fondazione ICSC, Spoke 3 Astrophysics and Cosmos Observations. National Recovery and Resilience Plan (Piano Nazionale di Ripresa e Resilienza, PNRR) Project ID CN\_00000013 "Italian Research Center on High-Performance Computing, Big Data and Quantum Computing" funded by MUR Missione 4 Componente 2 Investimento 1.4: Potenziamento strutture di ricerca e creazione di "campioni nazionali di R\&S (M4C2-19)" - Next Generation EU (NGEU).

\end{credits}
%\bibliographystyle{splncs04}
%\bibliography{references}

\begin{thebibliography}{1}
\providecommand{\url}[1]{\texttt{#1}}
\providecommand{\urlprefix}{URL }
\providecommand{\doi}[1]{https://doi.org/#1}

\bibitem{ahrens200536}
Ahrens, J., Geveci, B., Law, C., Hansen, C., Johnson, C.: 36-paraview: An
  end-user tool for large-data visualization. The visualization handbook
  \textbf{717},  50038--1 (2005)

\bibitem{adass_vlkb}
{Butora}, R., {Molinaro}, M.: {Interoperable standardisation of the VLKB
  dataset access services}. In: {Hugo}, B.V., {Van Rooyen}, R., {Smirnov}, O.M.
  (eds.) Astromical Data Analysis Software and Systems XXXI. Astronomical
  Society of the Pacific Conference Series, vol.~535, p.~299 (May 2024)

\bibitem{adass_vlw}
{Tudisco}, G., {Vitello}, F., {Sciacca}, E., {Riggi}, S., {Molinari}, S.,
  {Malikova}, E., {Krokos}, M.: {ViaLactea: a distributed Visual Analytic
  system for exploring our Galactic ecosystem}. In: {Hugo}, B.V., {Van Rooyen},
  R., {Smirnov}, O.M. (eds.) Astromical Data Analysis Software and Systems
  XXXI. Astronomical Society of the Pacific Conference Series, vol.~535, p.~211
  (May 2024)

\bibitem{Vitello2018}
Vitello, F., et~al.: Vialactea visual analytics tool for star formation studies
  of the galactic plane. Publications of the Astronomical Society of the
  Pacific  \textbf{130}(990),  084503 (jun 2018).
  \doi{10.1088/1538-3873/aac5d2}

\end{thebibliography}

\end{document}